\documentclass[a4paper,11pt]{article}
\usepackage{jheppub} 
\usepackage[T1]{fontenc} 
\RequirePackage{graphicx}
\usepackage{enumerate}
\usepackage{bm}
\usepackage{braket}
\usepackage{slashed}
\usepackage{comment}
\usepackage{color}

\title{The Exact WKB analysis and the Stokes phenomena of the Unruh
effect and Hawking radiation}
\author[a]{Seishi Enomoto}
\affiliation[a]{School of Physics, Sun Yat-sen University, Guangzhou 510275, China}
\author[b]{Tomohiro Matsuda}
\affiliation[b]{Laboratory of Physics, Saitama Institute of Technology,
Fukaya, Saitama 369-0293, Japan}
\emailAdd{seishi@mail.sysu.edu.cn}
\emailAdd{matsuda@sit.ac.jp}

\abstract{
The physical observables of quantum theory can be described by
perturbation theory, which is often given by diverging power series. 
This divergence is connected to the existence of non-perturbative
phenomena, where resurgence allows us to study this connection.
Applying this idea to the WKB expansion, the exact WKB analysis gives
a clear connection to non-perturbative phenomena.
In this paper, we apply the exact WKB analysis to the Unruh effect and
Hawking radiation.
The mechanism we found in this paper is similar to the Schwinger effect
of a constant electric 
field, where the background is static but the Stokes
phenomenon appears in the temporal part.
Comparing this with a sonic black hole, our calculations show a clear
discrepancy between them.
Then, we briefly explain how quantum backreactions can be included 
 in the exact WKB formalism. } 
\begin{document}

\maketitle
\section{Introduction}
\label{sec-intro}
The Stokes phenomenon is a very important phenomenon that directly
explains non-perturbative particle production.
It has been studied in a very broad range of applications. 
If special functions are used for the solutions, the Stokes phenomenon
is understood as a fundamental property of the special functions. 
Since the Stokes phenomenon is like a fingerprint, it can be used
effectively to distinguish phenomena that appear to be identical.  
When discussing a model in which particles are generated from a vacuum,
it is very useful to focus on the Stokes phenomenon.

The Schwinger effect is a very well-known model of non-perturbative
particle production and has already been studied in detail. 

If one studies Hawking radiation after learning about the Schwinger
effect, one will notice that many textbooks mention the similarity
between the Schwinger effect and Hawking radiation. 
Some may be tempted to try the same analytical methods that were
successful with the Schwinger effect. 
Some may try to figure out the difference between the Schwinger effect
and the Hawking radiation by comparing their Stokes phenomena.
However, such people will be puzzled there, since 
no article or textbook describes the Stokes phenomenon of Hawking
radiation as it appears in the Schwinger effect. 

Some may try to find a way by making use of Hawking's tunneling
hypothesis. 
This is because tunneling phenomena in quantum theory can generally be
described by the Stokes phenomenon. 
But here, too, one encounters a strange situation.
The analysis using the tunneling hypothesis is about ``phenomena that do not
generate the Stokes phenomena''(with the exception of some
analogues, which will be described later.) 

To the best of our knowledge, the Stokes phenomenon directly
related to radiation near the horizon has been an unsolved problem for
about half a century since Hawking's paper.
(As we will mention later in this paper, the Stokes phenomena away from
the horizon are already known and has been used in the analysis of the
glaybody factor and quasinormal modes.) 

In this section, we will explain the above questions in some detail.
Once the details of the Stokes phenomena are known, the distinction
from black hole analogues will become clearer.
If the Stokes phenomenon is analyzed with the Exact WKB, quantum
reactions can be treated more rigorously.  
We will discuss these issues in the latter part of this paper.

The tunneling problem in quantum mechanics is distinct from that in
classical physics. 
The problem is usually discussed for quantum particles that pass through
a potential barrier, which is forbidden in classical mechanics. 
The Wentzel Kramers Brillouin Jeffreys (WKB or WKBJ) approximation, a
method for finding approximate solutions to linear differential
equations, is considered the most useful for this problem \cite{Berry:1972na,Landau-Lifshitz}.
The WKB approximation has been applied to quantum fields propagating in
curved backgrounds, such as Hawking evaporating
processes \cite{Kraus:1994fj,Parikh:1999mf, Keski-Vakkuri:1996wom,
Srinivasan:1998ty, Shankaranarayanan:2000qv,
Banerjee:2009wb,Arzano:2005rs, Akhmedova:2008dz, Singleton:2013,
Dumlu:2020wvd} and inflationary cosmological 
perturbations \cite{Martin:2002vn}.
It has also been widely used to determine quasi-normal modes of black
holes \cite{Iyer:1986np, Konoplya:2003ii, Konoplya:2011qq}.

For a particle with energy $E$ and rest mass $m$ moves in a
one-dimensional potential $V(x)$ with $E<V(x)$ in the range 
$a\ge x \ge b$, the amplitude of the tunneling is described by
\begin{eqnarray}
\label{eq-wkb0}
e^{-\frac{1}{\hbar}\int^b_a\sqrt{2m\left(V(x)-E\right)}dx}
&=&e^{-\frac{1}{\hbar}\int^b_a p(x)dx},
\end{eqnarray}
where $p(x)$ denotes the canonical momentum.
If $V(x)$ is given by an inverted quadratic potential, exact solutions
described by special functions can be obtained.
It is also known that the local analysis near a turning point is
well described by the Airy function \cite{Berry:1972na,Landau-Lifshitz}.

For $E>V(x)$, a similar quantum effect can be seen in the reflection,
since the reflection of the particle is quantum \cite{Berry:1972na,Landau-Lifshitz}. 
Describing this phenomenon using Eq.(\ref{eq-wkb0}) is a good idea in
practice, but in this case, the situation is less obvious \cite{Berry:1972na}. 
First, the turning points defined by $p(x)=0$ are not on the real axis
of $x$. 
In the simplest case of an inverted quadratic potential, the pair of
turning points appears on the imaginary axis of $x$. 
In this model, one can find exact solutions described by
special functions (e.g., Weber functions or parabolic cylinder
functions).

In more general situations, the Exact WKB(EWKB) formalism is useful.
Initially, the formalism was valid only for second-order equations,
but later, with the discovery of new Stokes lines \cite{NBR} and
virtual turning points \cite{Virtual:2015HKT},
 it was found to be valid for higher-order differential equations.
For more mathematical details on resurgence and the EWKB formalism,
see Refs.\cite{RPN:2017, Virtual:2015HKT, Voros:1983, Delabaere:1993,
Silverstone:2008, Aoki:2009, Takei:2008}.
Using the Borel resummation, the WKB expansion is cast into a 
so-called Borel panel, where finite and exact results are extracted.
The EWKB consists of microlocal reduction of ordinary differential
operators and its application to global analysis.

In resurgence, non-perturbative effects appear from divergence.
On the other hand, WKB approximation has to discard higher terms,
from which the divergence appears.
The EWKB helps us to understand why fine results are computed using such
an approximation.
Without the EWKB, it is also difficult to understand how higher terms
affect the Stokes lines and their Stokes phenomena. 
Note that the definition of turning points is also less rigorous in the
WKB approximation when compared to the EWKB;
the EWKB is exact about the $\hbar$ expansion, which removes such
ambiguity.
In physics, the EWKB gives connection formulae, which are used
to calculate the Bogoliubov coefficients of quantum particle
production \cite{Kofman:1997yn,Enomoto:2020xlf} and quantum transitions
in the Landau-Zener model \cite{Zener:1932ws,Enomoto:2021hfv};
for recent applications of the EWKB, see Ref.\cite{WKB-recent:2022}.

In this paper, the Stokes phenomenon for second-order ordinary
differential equations is considered in the framework of the WKB
formalism. 
The physical observables of quantum theory can be described by
perturbation theory (WKB expansion), which often gives a diverging power
series. 
This divergence is related to the existence of non-perturbative
phenomena (Stokes phenomena), so the definition of the vacuum by the WKB
approximation becomes ambiguous for non-perturbative phenomena; using
EWKB, definition of the vacuum becomes rigorous.

Since the Stokes phenomenon does not occur within the Stokes domain
 (the domain bounded by the Stokes lines),
we need to find the Stokes lines where (or when) the particles are
created.
On the other hand, if the definition of the vacuum is independent of the
observer's differential equation, the observer may see radiation because
the two sets of vacuum definitions are not identical.
Both the Unruh effect \cite{Unruh:1976db} and 
Hawking radiation \cite{Hawking:1975vcx} have a static curved
 background, but we found the Stokes phenomenon in the temporal part.
Our calculation uses the discrepancy of the vacua 
and the Stokes phenomenon at the same time.

Applying the EWKB to quantum scattering due to an inverted
quadratic potential with $E>V(x)$, the Stokes lines is a Merged pair of
Turning Points (MTP), whose exact solutions are given by the Weber function;
details are described in Sec.\ref{sec-revWKB}.
The MTP structure often appears in non-perturbative analysis 
where a local linear approximation is valid, such as 
particle production in standard preheating
scenarios \cite{Kofman:1997yn, Enomoto:2020xlf} and the Landau-Zener
transition \cite{Zener:1932ws,Enomoto:2021hfv}.
In these models, the $x$-dependence is replaced by a $t$-dependence, 
and local linear approximations are found in the
scalar mass ($m_\varphi(t)\simeq vt$) of the preheating scenario and
in the linear diagonal elements ($D_{\pm}(t)\simeq \pm vt$) of 
the Landau-Zener model \cite{Enomoto:2022nuj}.

When the differential equations have poles in addition to 
turning points, the situation becomes less straightforward, but 
the Stokes phenomenon can still be described in the EWKB formalism
\cite{Koike:2000}.
In the space part of the WKB formalism, the Stokes lines of 
the Unruh effect and Hawking radiation typically take a loop structure
around a pole at the horizon, where the Stokes lines do not exist in the
vicinity \cite{Dumlu:2020wvd}. 
Since there is no Stokes line near the horizon, it is not possible
to see the mixing of $\pm$ solutions near the horizon (at least in the
spatial part).
On the other hand, for an observer standing far from the black hole,
the Stokes phenomenon may appear midway between the observer and the
horizon.
These Stokes phenomena have been used to calculate black hole 
excitations and greybody factors \cite{Dumlu:2020wvd}, 
consistent with other calculations \cite{Maldacena:1997ih}.
To find discrepancies between the definition of the inertial vacuum and
the observer's vacuum, global analyses are often used, in which global
analytic conditions on the complex plane of the given 
coordinate system are examined \cite{Birrell:1982ix,Frolov:1998wf}. 

Since the Unruh effect (and Hawking radiation) are intuitively local events, 
it seems quite natural to expect that the effects extracted from the
global analysis would be encoded in the local coordinate system.
An analogy with the Schwinger effect would be useful to understand 
the effects in the local coordinate system.
It is already known \cite{Frolov:1998wf} that the Unruh effect is very
similar to the Schwinger effect, but the corresponding Stokes phenomenon
has not been found in the Unruh effect;
using the EWKB approach, we find the Stokes phenomenon in the time
part of the WKB formalism of the Unruh effect.
As described in Sec. 10.2 of Ref.\cite{Frolov:1998wf}, a naive
application of the Schwinger effect to the Unruh effect (Hawking
radiation) simply replaces the field strength with the proper
acceleration (surface gravity) but requires a factor of 2 
($\beta=2$ in Ref.\cite{Frolov:1998wf}).
The present calculation reproduces the factor two in the Unruh effect,
but no such factor is required in Hawking radiation.

To compare the WKB approximation and the EWKB, we begin with
 the simplest calculation of the steepest descent path method 
of the WKB approximation.\footnote{This calculation
 should not be confused with the exact calculation described 
in Ref.\cite{Aokisteepest:2004}.}
The method of steepest descent path was first applied to cosmological
particle production in Ref.~\cite{Chung:1998bt} and extended for higher
polynomials in Ref.\cite{Enomoto:2013mla, Enomoto:2014hza}.
Consider an equation of motion given by
\begin{equation}
 \ddot{\chi}_k+\left[k^2+m_0^2-V(t)\right]\chi_k = 0,
  \label{eq:eom_chi_k_2}
\end{equation}
where one can choose $V(t)=-\frac{g_{2n}^2}{M_*^{2(n-1)}}(vt)^{2n}$.
(Here we temporarily assume $\hbar=1$ and $c=1$.)
For $n=1$, this equation coincides with the original preheating scenario
of Ref.\cite{Kofman:1997yn} and is similar to the scattering problem by an
inverted quadratic potential.
One can write down the (conventional) WKB
solution given by
\begin{equation}
 \chi_k = \frac{\alpha_k}{\sqrt{2\omega_k}} e^{-i\int_{-\infty}^t dt'\omega_k(t')}
   + \frac{\beta_k}{\sqrt{2\omega_k}} e^{+i\int_{-\infty}^t
   dt'\omega_k(t')}, \label{eq:WKB-type_solution}
\end{equation}
where
\begin{equation}
 \omega_k\equiv \sqrt{k^2+m_0^2-V(t)}
\end{equation}
and $\alpha_k, \beta_k$ have to satisfy
\begin{equation}
 |\alpha_k|^2 - |\beta_k|^2 = 1.
\end{equation}
We consider the additional constraint (the calculation corresponds to
the simplest choice of parameters in Ref.\cite{Dabrowski:2016tsx})
\begin{equation}
 0 = \dot{\left(\frac{\alpha_k}{\sqrt{2\omega_k}}\right)} e^{-i\int_{-\infty}^t dt'\omega_k(t')}
   + \dot{\left(\frac{\beta_k}{\sqrt{2\omega_k}}\right)} e^{+i\int_{-\infty}^t
   dt'\omega_k(t')}.
\end{equation}
Substituting the representation (\ref{eq:WKB-type_solution}) into
(\ref{eq:eom_chi_k_2}) with the above constraint, one obtains 
\begin{eqnarray}
 \dot{\alpha}_k &=& \beta_k \frac{\dot{\omega}_k}{2\omega_k} e^{+2i\int_{-\infty}^t dt' \omega_k},\\
 \dot{\beta}_k &=& \alpha_k \frac{\dot{\omega}_k}{2\omega_k}
  e^{-2i\int_{-\infty}^t dt' \omega_k}. \label{eq:eom_beta}
\end{eqnarray}
Supposing that the initial state has no particle, we have 
\begin{equation}
 \alpha_k(-\infty)=1, \qquad \beta_k(-\infty)=0.
\end{equation}
One can estimate the Bogoliubov
coefficients for $\beta_k \ll 1$ by the integration given by
\begin{equation}
\label{eq-bogint}
 \beta_k(+\infty) = \int_{-\infty}^{+\infty} dt  \frac{\dot{\omega}_k(t)}{2\omega_k(t)}
  \exp{\left[ -2i \int_{-\infty}^t dt' \omega_k(t') \right]}.
\end{equation}
This corresponds to the so-called instantaneous Hamiltonian
diagonalization method \cite{Haro:2010mx}.
The factor $1/\omega_k$ at $\omega_k(t)=0$ of the $t$-integral
is important for the calculation in Ref.\cite{Chung:1998bt}.
The integral in Eq.(\ref{eq-bogint}) can be evaluated on the complex $t$ plane
using the steepest descent method.
For details of the calculation, see Ref.\cite{Chung:1998bt,
Enomoto:2013mla, Enomoto:2014hza}.
Although the above formulation is very simple and convenient for numerical calculations, 
it does not account for resurgence.
In fact, although resurgence is known to link the divergence of the WKB
approximation to non-perturbative phenomena, such higher-order terms are
discarded from the outset in the above calculation.
This discrepancy is not trivial.

The calculation of the Unruh and Hawking effects by the
WKB tunneling method is expected to explain the heuristic picture of
Hawking described in Ref.\cite{Hawking:1975vcx}.
Hawking explains the effect as a tunneling outward of positive energy
modes and a tunneling inward of negative energy modes.
This heuristic picture has been explored later in
Ref.\cite{Kraus:1994fj, Srinivasan:1998ty, Parikh:1999mf, 
Shankaranarayanan:2000qv, Banerjee:2009wb, Arzano:2005rs, Akhmedova:2008dz,
Singleton:2013, Dumlu:2020wvd, Keski-Vakkuri:1996wom}, in which 
the action of a particle picks up an imaginary contribution when it
crosses the horizon (the integral path is taken on the complex plane),
 and the imaginary contribution was interpreted as tunneling probability.
When the static black hole is viewed in Kruskal coordinates 
 valid on both sides of the horizon, the factor computed by
``tunneling''\footnote{Note however that normally the modes have $E>V(r)$ near
the horizon.} 
is nothing more than a factor necessary for the WKB
approximate solutions to connect smoothly inside and outside the
horizon \cite{Srinivasan:1998ty, Banerjee:2009wb}, and the Stokes
phenomena do not appear in the computation. 
In fact, since the EWKB solutions of the spatial part are in the same
Stokes domain, the Stokes phenomenon does not appear by the ``tunneling''.

For an analog black hole (sonic black hole),
Giovanazzi \cite{Giovanazzi:2004zv} found that a microscopic description of Hawking
radiation explains Hawking's original picture of tunneling.
Giovanazzi's model explains how the Stokes phenomenon can be
responsible for Hawking radiation in the analog system.
These analog systems are expected to gain a better understanding of
Hawking radiation avoiding theoretical problems of the original
black hole radiation.
However, comparing the Stokes phenomena of these systems, there seems to
be a crucial discrepancy between them.
In Sec.\ref{sec-revWKB}, we will show some details of the Stokes
phenomena to compare the mechanisms of these models. 

Sec.\ref{sec-qb} briefly describes the Stokes phenomena of
Hawking radiation and a sonic black hole when quantum effects are
involved, and shows why the exact definition of the
$\hbar$-expansion in the EWKB is important for understanding physics beyond
the semiclassical approximations.

\section{The EWKB for the Stokes phenomena and Hawking radiation}
\label{sec-revWKB}
The mathematical formulation of the EWKB uses $\eta\equiv \hbar^{-1}\gg 1$ for
the expansion, instead of using the Planck (Dirac) constant $\hbar$.
Following Ref.\cite{Voros:1983, Virtual:2015HKT}, we start with the
``Schr\"odinger equation'' in quantum mechanics described 
as\footnote{Note that to avoid confusion, the second derivative of the 
Schr\"odinger equation should be preceded by $1/2m$.
Here we follow the standard mathematician's formulation of the EWKB.
The above ``Schr\"odinger equation'' is not identical to the
Schr\"odinger equation in physics.}
\begin{eqnarray}
\left[-\frac{d^2}{dx^2}+\eta^2 Q(x)
\right]\psi(x,\eta)&=&0.
\end{eqnarray}
Introducing the ``potential'' $V(x)$ and the ``energy'' $E$, we define
\begin{eqnarray}
Q(x)&=&V(x)-E.
\end{eqnarray}
Although an explicit $\eta$-dependence is not required in this section, 
 for our later argument we write it in the following form;
\begin{eqnarray}
Q(x,\eta)&=&\sum^\infty_{j=0} \eta^{-j} Q_j(x).
\end{eqnarray}
Writing a WKB solution as $\psi(x,\eta)=e^{R(x,\eta)}$,
we have for $S(x,\eta)\equiv \partial R/\partial x$,
\begin{eqnarray}
\psi&=&e^{\int^x_{x_0}S(x,\eta)dx},
\end{eqnarray}
where $x_0$ is normally taken at a turning point of the Stokes domain.
For $S$, we have 
\begin{eqnarray}
-\left(S^2 +\frac{\partial S}{\partial x}\right)+\eta^2 Q&=&0.
\end{eqnarray}
If one expands $S$ as $S(x,\eta)=\sum_{n=-1}^{n=\infty}\eta^{-n} S_{n}$,
one will find
\begin{eqnarray}
S=\eta S_{-1}(x)+ S_0(x)+\eta^{-1}S_1(x)+...,
\end{eqnarray}
which leads to the following recursive relations
\begin{eqnarray}
\label{eq_firstterm}
S_{-1}^2&=&Q_0\\
2S_{-1}S_{j+1}&=&-\left[\sum_{k=0}^jS_kS_{j-k}
+ \frac{d S_{j}}{dx} -Q_{j+2}\right].
\end{eqnarray}
Points of $Q_0(x)=0$ (which is not always identical to $Q(x,\eta)=0$)
are called turning points of the equation.\footnote{
This definition of the turning point differs from the definition usually
used in physics.}
See Ref.\cite{Aoki:1993} for details of the calculation when $Q(x,\eta)$
depends explicitly on $\hbar=\eta^{-1}$.
Later we will consider $Q(x,\hbar)$ when quantum effects are involved, but
here we temporarily assume $Q(x,\eta)=Q_0$ .
 After some calculation, one will have 
\begin{eqnarray}
\label{eq-sodd}
\psi&=&\frac{1}{\sqrt{S_{odd}}}e^{\int^x_{x_0}S_{odd}dx}\\
&&S_{odd}\equiv\sum_{j\ge 0}\eta^{1-2j}S_{2j-1}.
\end{eqnarray}
Depending on the sign of the first $S_{-1}=\pm \sqrt{Q(x)}$, there are
two solutions $\psi_\pm$, which are given by
\begin{eqnarray}
\psi_{\pm}&=&\frac{1}{\sqrt{S_{odd}}}\exp\left(\pm \int^x_{x_0}S_{odd}
					  dx\right)\nonumber\\
&=&e^{\pm \eta \int\sqrt{Q}dx}\sum^\infty_{n=0}\eta^{-n-1/2}\psi_{\pm,n}(x).
\end{eqnarray}
The above WKB expansion is usually divergent but is Borel-summable.
The Borel transform is taken as
\begin{eqnarray}
\psi_{\pm}^B&=&\sum^\infty_{n=0}
\frac{\psi_{\pm,n}(x)}{\Gamma(n+\frac{1}{2})}\left(y\pm 
 s(x)\right)^{n-\frac{1}{2}}.
\end{eqnarray}
Theorems and proofs of the EWKB are mostly given on the
``Borel panel'' using $\psi^B_\pm$.
For physics, what is important here is the shift of the integral
 of the inverse-Laplace integration.
The shift is determined by $S_{-1}$ as
\begin{eqnarray}
\psi_\pm &\rightarrow&\Psi_\pm\equiv\int^\infty_{\mp
 s(x)}e^{-y\eta}\psi_\pm^B(x,y)dy,\\
\label{eq_sx}
&&s(x)\equiv \int^x_{x_0}S_{-1}(x)dx,
\end{eqnarray}
where the $y$-integral is parallel to the real axis.
The Stokes lines are defined as the solutions of $\mathrm{Im} [s(x)]=0$.
In the EWKB, the Stokes lines are exact with respect to the $\hbar$
expansion, and they are computed from first-order terms.
This is a striking result given that the Stokes lines determine the 
intrinsic properties of quantum systems;
without the EWKB, it is difficult (or impossible) to prove that 
the Stokes lines are unaffected by higher order terms(e.g, $S_0,S_1, ...$). 
(We are not showing proofs here. See \cite{RPN:2017, Delabaere:1993,
Virtual:2015HKT} for more details.)
Such higher-order terms are often used to avoid
discontinuities in the Stokes phenomena in semiclassical
approximation \cite{Dabrowski:2016tsx}, but as long as the WKB
approximation is used, the dilemma was that the inclusion of such terms
could easily change the structure of the Stokes lines.
The Exact WKB removes the notorious uncertainty of the WKB
approximation and redefines the semiclassical calculation.

In typical cases, the Stokes lines are often degenerate.
In such cases, a gap may appear at $\mathrm{Im}[\eta]=0$ (between $\mathrm{Im}[\eta]>0$ and
$\mathrm{Im}[\eta]<0$).
This gap has already been resolved analytically by the normalization
factor (called Voros coefficient \cite{Voros:1983}), whose analytic
calculation is given in Ref.\cite{Voros:1983,Silverstone:2008, Takei:2008, Aoki:2009} for MTP
(Merged pair of simple Turning Points) and a loop structure of a
Bessel-like equation \cite{Aoki:2019}.
More simply, the normalization coefficient
can be computed using a simple consistency relation \cite{Berry:1972na}, 
but that method cannot determine the phase of the coefficient.

To explicitly show the calculation of the connection matrix using the
EWKB, we begin with\footnote{This equation appears in the cosmological preheating
scenario of Ref.\cite{Kofman:1997yn}.}
\begin{eqnarray}
\frac{d^2 \psi}{dt^2}+\eta^2\left[E+g^2_2v^2t^2\right]\psi=0. \label{eq:eom_inverted_potential}
\end{eqnarray}
For $E<0$, turning points will appear on the real axis.
This is a conventional tunneling problem of quantum mechanics.

\begin{figure}[t]
\centering
\includegraphics[width=0.5\columnwidth]{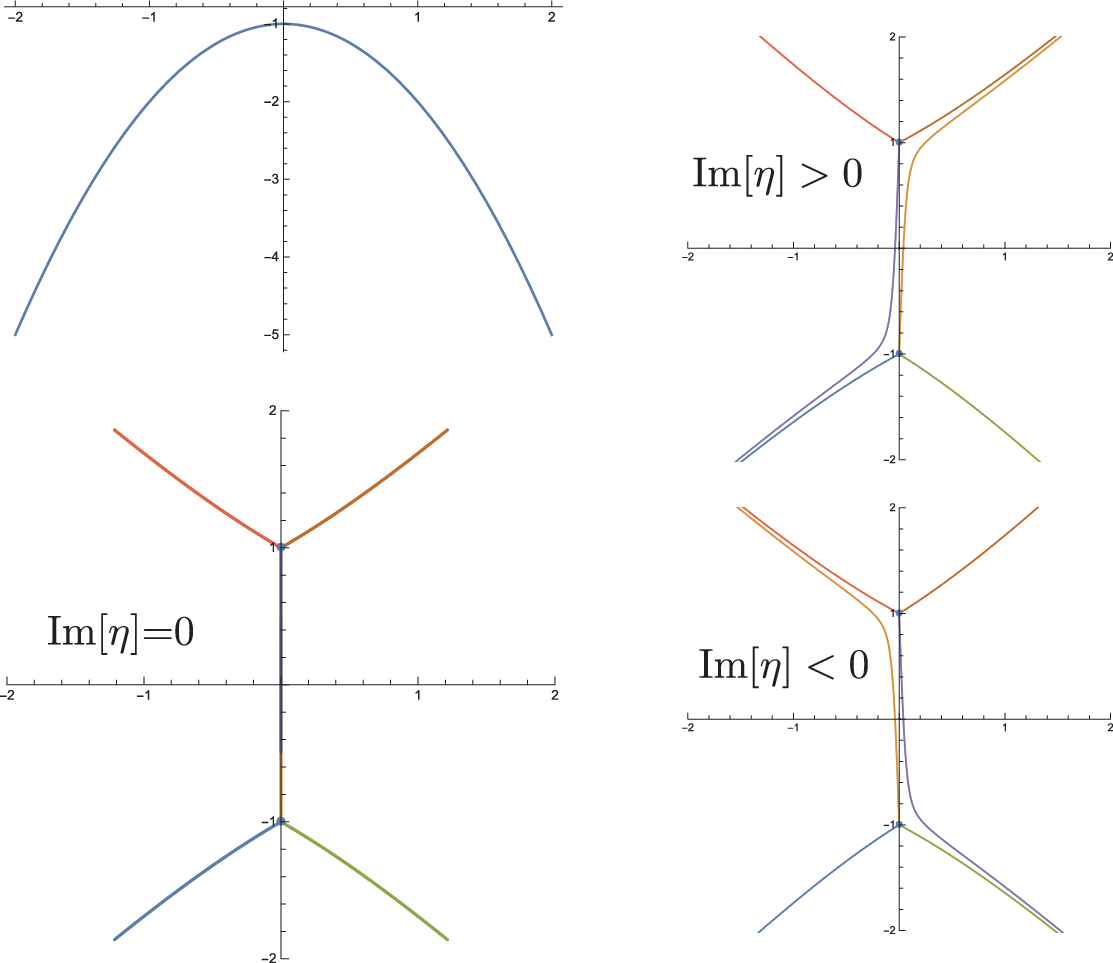}
 \caption{The potential $V(t)=-1-t^2$ ($E=0$) is shown on the upper left
 panel, in which the horizontal axis is real $t$. 
Other panels are the Stokes lines for Im$[\eta]=0$ (lower left),
 Im$[\eta]>0$(upper right) and Im$[\eta]<0$(lower right) shown 
on the complex  $t$-plane.
The Stokes lines have degenerated for Im$[\eta]=0$ and
 there is a gap between Im$[\eta]>0$ and Im$[\eta]<0$.
The MTP structure has turning points at $t=\pm i$.}
\label{fig-quadscat-stokes}
\end{figure}

For $E>0$, turning points will appear on the imaginary $t$-axis.
In Fig.\ref{fig-quadscat-stokes}, we show the Stokes lines for $V(x)=-1-t^2$ ($E=0$).
Assuming $E=k^2>0$, 
the connection matrix starting from $t=-\infty$ to $t=+\infty$ can be
calculated using the EWKB, which gives
(See Ref.\cite{Enomoto:2020xlf} for calculational details)
\begin{eqnarray}
\left(
\begin{array}{cc}
N^2 \left(e^{-K_{ud}} +e^{K_{ud}} \right) & -i e^{-K_{ud}} \\
 i e^{-K_{ud}} & \frac{e^{-K_{ud}}}{N^2} \\
\end{array}
\right),
\end{eqnarray}
where we defined
\begin{eqnarray}
\label{eq-kud}
K_{ud}&\equiv&  \int^{t_*^-}_{t_*^+}S_{odd}dt \label{eq:K_ud}
\end{eqnarray}
for $t_*^\pm\equiv \pm i\frac{k}{g_2v}$.
To make the diagonal elements consistent, one may put a condition on the
normalization factor $N$;
\begin{eqnarray}
|N|^{-2}=\sqrt{1+e^{2K_{ud}}}.
\end{eqnarray}
Finally, the connection matrix becomes
\begin{eqnarray}
\left(
\begin{array}{cc}
\sqrt{1+e^{-2K_{ud}}} & -i e^{-K_{ud}} \\
 i e^{-K_{ud}} & \sqrt{1+e^{-2K_{ud}}} \\
\end{array}
\right).
\end{eqnarray}
The result is consistent with the exact result obtained from the Weber
function \cite{Enomoto:2020xlf}. 

The Schwinger effect can also be discussed in the same
way \cite{Schwinger:1951nm, Haro:2010mx, Taya:2020dco, Kitamoto:2020tjm,
Martin:2007bw, Dumlu:2010ua}.
Consider a complex charged scalar field interacting with an
electromagnetic field with the vector potential (constant electric field
$E_z=E$)
\begin{eqnarray}
A_\mu(t)&=&(0,0,0,-Et),
\end{eqnarray}
one will have
\begin{eqnarray}
\hbar ^2\frac{d^2 \phi_\mathbf{k}}{c^2dt^2}+\left[m^2 c^2+\left(
\mathbf{p}-\frac{q}{c}\mathbf{A}(t)\right)^2
\right]\phi_\mathbf{k}&=&0,
\end{eqnarray}
which gives for the EWKB,
\begin{eqnarray}
Q(t)&=&-m^2c^4-
c^2\left(
\mathbf{p}-\frac{q}{c}\mathbf{A}(t)\right)^2.
\end{eqnarray}
Using the EWKB, the Stokes lines can be easily drawn (changing the
variable result in a simple form \cite{Haro:2010mx}) and the connection
matrix of the solutions can be calculated. 
See also the exact solution given by the parabolic cylinder functions 
in Ref.\cite{Haro:2010mx}.
This situation is exactly the same as the cosmological particle
production in the preheating model \cite{Kofman:1997yn}.
In the above equation, particle production appears to be a local event at a
 specific time, but given gauge symmetry, the time is arbitrary as long
 as the electric field is constant.
See also Ref.\cite{Dumlu:2010ua} for a discussion of the Stokes
phenomena and Schwinger pair production in time-dependent laser pulses.
Given the similarity between the Schwinger effect and Hawking
 radiation \cite{Frolov:1998wf}, it seems quite natural that the
 Stokes phenomenon in Hawking radiation must be explained by a similar
 Stokes phenomenon in the temporal part.
However, to the best of our knowledge, such a Stokes phenomenon has not
been discussed yet in past studies.
In this paper, we will show how the Stokes phenomenon appears in the Unruh
effect and apply it to Hawking radiation.

\subsection{The Stokes phenomena of the Unruh effect}
\begin{figure}[t]
\centering
\includegraphics[width=0.5\columnwidth]{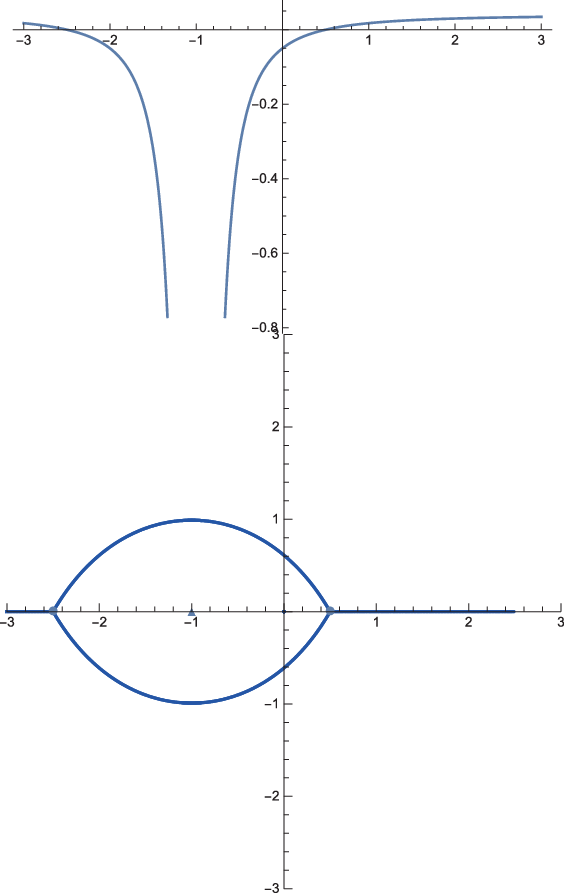}
 \caption{The upper panel is the ``potential'' of
 $Q(x_R)=-\frac{E^2-m^2(1+ax_R)^2}{(1+ax_R)^2}$ for
 $a=1,E=0.3,m= 0.2$ plotted for real $x_R$ (horizontal axis),
 and the lower panel shows the Stokes lines on the complex $x_R$ plane.
A pole is shown at the center of the loop.}
\label{fig-Rindler-x}
\end{figure}

Let us take a quick look at how the Stokes lines of the Unruh effect look 
in Rindler coordinates;
in the WKB formalism, the spatial and temporal part can be
separated.
Here we consider a two-dimensional
Rindler metric \cite{Birrell:1982ix}
\begin{eqnarray}
ds^2&=&-\left(1+\frac{a x_R}{c^2}\right)^2 c^2dt_R^2+dx_R^2.
\end{eqnarray}
Assuming $c=1$, corresponding EWKB formulation has the space part given 
by\cite{Singleton:2013}
\begin{eqnarray}
 Q(x_R)=-\frac{E^2-m^2 (1+ax_R)^2}{(1+ax_R)^2},
\end{eqnarray}
where two turning points and a pole appear at
$x_R=(\pm E/m-1)/a$ and at the midpoint $x_R=-1/a$ (at the 
center of the loop structure of the Stokes lines in
Fig.\ref{fig-Rindler-x}).
($E$ in the scalar field equation must be discriminated form $E$ in
the ``Schwinger equation''.)
In the $m\rightarrow 0$ limit, these turning points will flow to infinity.
Since no Stokes line exists within the loop structure, the Stokes
phenomenon is not expected in the space part.
The ``potential'' and the Stokes lines are shown in
Fig.\ref{fig-Rindler-x}.
Although there is no Stokes phenomenon in the space part of $x_R$, 
the ``tunneling amplitude'' at the pole (on the horizon)
 introduces  $e^{- \pi \omega/2a}$.
As discussed in Ref.\cite{Akhmedova:2008dz, Singleton:2013},
the Unruh effect is accounted for using the WKB/tunneling method,
 but this calculation had a factor two problem.
In Ref.\cite{Akhmedova:2008dz,Singleton:2013}, a resolution of this
discrepancy was advocated, in which authors pointed out that a time part
contribution gives 
Im$[\omega \Delta t]=-\pi\omega/a$ upon crossing the horizon, where the
time and the space coordinates reverse their time-like/space-like
character.
Using them as global analytic conditions \cite{Birrell:1982ix,
Frolov:1998wf, Nakamura:2007vp} and comparing the modes of inertial vacuum
with the modes of an accelerating observer, one will find the Unruh effect. 

Since the Unruh effect is a local event, it is natural to expect that 
its information must be encoded in the local coordinate system.
To understand what the observer sees in the inertial vacuum,
we write a pair of $e^{\pm i \int \omega dt}$ solutions of the inertial
observer's vacuum using the accelerating observer's time ($\tau$).
The point of our observation is that $t$ is curved with respect to
$\tau$, and such curvature can often introduce the Stokes phenomena.

Considering $ct=\frac{c^2}{a}\sinh \left(\frac{a}{c} \tau\right)$, 
we find $dt= \cosh\left(\frac{a}{c} \tau\right) d\tau$.
Using $\alpha\equiv a/c$, we have
\begin{eqnarray}
\label{eq-stokes1}
e^{\pm i \int \omega dt}&=&
e^{\pm i \int \omega \cosh(\alpha \tau) d\tau},
\end{eqnarray}
which corresponds to the EWKB formalism with
\begin{eqnarray}
\eta^2 Q(\tau)&=& - \omega^2  \cosh^2 (\alpha \tau).
\end{eqnarray}
Here we have skipped the
formal WKB expansion. Formally, one has to solve the differential
equation using the WKB expansion. The equation is trivial for $t$ but it
becomes non-trivial after changing the variable from $t$ to $\tau$.
Considering a local series expansion of $Q(\tau)$ around $\tau=0$
\begin{eqnarray}
\eta^2Q(\tau)&=& - \omega^2  \cosh^2 (\alpha \tau)\nonumber\\
&\simeq&-\omega^2-\alpha^2 \omega^2\tau^2,
\end{eqnarray}
one can expect that the Stokes lines will form an MTP structure around
$\tau=0$.
The Stokes phenomenon of the MTP structure gives mixing of $\pm t$
(i.e, $\pm\tau$) solutions with $|e^{K_{ud}}|^2=e^{- \pi \omega/\alpha}$.
Note that in the local coordinate system around $\tau=0$, we also have
the expansion
\begin{eqnarray}
\omega t&=&\omega \left(\frac{1}{\alpha}\sinh (\alpha \tau)\right)\simeq \omega \tau.
\end{eqnarray}
See also the remark on page 47 of Ref.\cite{Birrell:1982ix}.

\begin{figure}[t]
\centering
\includegraphics[width=0.5\columnwidth]{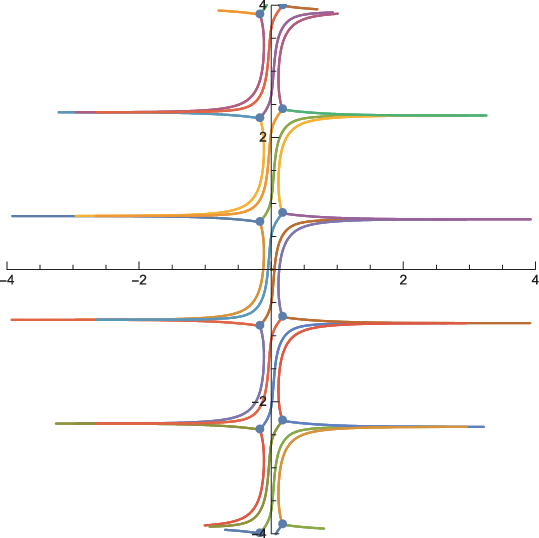}
 \caption{The Stokes lines for $Q(\tau)=-(\cosh^2 2 \tau+0.1+0.1
 i)(1+0.1i)$ are shown on the complex $\tau$ plane.
The double turning points and the degenerated Stokes lines are both split by
 introducing small parameters.
Note however that imaginary $\eta$ alone cannot solve all the
 degeneracy of this model.}
\label{fig_cosh}
\end{figure}

The structure of the Stokes lines is shown in 
Fig.\ref{fig_cosh}, which is the MTP structure surrounded by
a pair of Stokes lines.
Because of the Lorentz symmetry, $\tau=0$ is not special as far as $a$ 
is constant. 

As we have mentioned previously, naive application of the Schwinger
effect to the Unruh effect requires a factor of two.
Our calculation of the local coordinate system gives 
$e^{- \pi\omega/\alpha}$ but the Unruh effect requires 
$e^{- 2\pi \omega/\alpha}$.
This indicates that the same problem occurs in our calculation.
This result is not surprising since both calculations are
local and the effect of the horizon is disregarded.
In the next section, we will look at some details of the argument
to see that this problem does not appear in Hawking radiation if the
Hawking's tunneling picture is properly incorporated.

\subsection{The Stokes phenomena of Hawking radiation}
We are going to see the Stokes phenomena of the modes that are defined
with respect to the Killing time but viewed from an observer fixed at
the horizon.
Then, the local coordinate system 
would exhibit the Stokes phenomenon similar to the Schwinger effect.
Despite their similarities, there is a very important difference
between them. 
If a pair of particles is on the same side of the horizon, their total energy
does not vanish, so this process is prohibited by the law of
conservation of energy.
Since the negative energy state must be located inside the horizon
and its pair state must be outside, the production of 
observable (not virtual) particles must occur very
close to the horizon (at least within Compton length of the particles), 
and only half of the created particles can be seen as 
``radiation from the horizon''.
Since the Killing horizon in static spacetime is known to be the event
horizon, (not virtual) particle production in a static gravitational system is
only possible at the event horizon.
In the case of black hole radiation, particle production must be pair
production, and only half of the particles are observed.
Then, the probability of particle production in the radiation must be
squared, and an additional factor of two is introduced into the
coefficient to obtain the correct Hawking temperature.

As mentioned earlier, the Stokes phenomena are not found in the
space part of the WKB formalism, but they are found in
the time part.
Our idea is simple and straight.
Since the Stokes phenomena of the Schwinger effect for a constant
electric field are found in the time part \cite{Haro:2010mx}, 
it is quite natural to expect that the Stokes phenomena of Hawking
radiation must also be found in the time part, even though both models
have ``static'' backgrounds.
Our speculation was correct.
The Stokes phenomena in the time part successfully explained Hawking radiation.

On the other hand, Giovanazzi \cite{Giovanazzi:2004zv} found
that a microscopic description of the space part of a sonic black hole 
explains the picture of ``Hawking radiation as the tunneling''.
Since Giovanazzi's explanation of the tunneling conjecture implicitly
uses the Stokes phenomena at the horizon, it must be important to compare these
models in light of the Stokes phenomena.
To explicitly show the difference between them, the next section
describes the Stokes phenomena of the sonic black hole in some detail.

\subsection{The Stokes phenomena of a sonic black hole}
\label{sec-sonic}
Considering a one-dimensional Fermi-degenerate liquid squeezed by a
smooth barrier forming a transonic flow,
Giovanazzi \cite{Giovanazzi:2004zv}
 presented a microscopic description of Hawking radiation in sonic black
 holes.

The fundamental equations of fluid dynamics for an inviscid fluid in one
dimension are the equation of continuity and Euler's equation given by
\begin{eqnarray}
\partial_t\rho\partial_x (\rho v)&=&0\nonumber\\
\rho(\partial_tv+v\partial_x v)&=&-\partial_x p-\frac{\rho}{m}\partial_x
 V(x),
\end{eqnarray}
where $\rho(t,x)$, $v(t,x)$, $p(t,x)$, $m$ and $V(x)$ are the density,
velocity of flow, pressure, mass of a particle, and external potential of
the particle, respectively.
Then expand $\rho$ and $v$ as
\begin{eqnarray}
\rho&=&\rho_0+\epsilon \rho_1+\epsilon^2 \rho_2+...,\nonumber\\
v&=&\rho_0+\epsilon v_1+\epsilon^2 v_2+...,
\end{eqnarray}
where $\rho_0$ and $v_0$ denote the background flow, and $\rho_i$ and
$v_i$ are giving the sound waves.
Introducing the velocity potential $\phi$ such that $v_1=-\partial_x
\phi$, one will find 
\begin{eqnarray}
\partial_\mu g^{\mu\nu}\partial_\nu \phi&=&0,
\end{eqnarray}
where $\mu,\nu =t,x$.
Here the ``acoustic metric'' is given by
\begin{eqnarray}
g_{\mu\nu}&=&\frac{c_s^2}{\rho_0^2}
\left(
\begin{array}{cc}
c_s^2-v_0^2 & v_0\\
v_0 & -1
\end{array}
\right),
\end{eqnarray}
where
\begin{eqnarray}
c_s&\equiv&\sqrt{\frac{\partial p}{\partial \rho}}.
\end{eqnarray}
Choosing an appropriate potential $V(x)$, one will find that Hawking
radiation is realized at the sound horizon which divides regions of
 $v_0<c_s$ and  $v_0>c_s$.
The temperature of the radiation at the horizon is
\begin{eqnarray}
T=\frac{\hbar c_s}{4\pi k_B}\frac{\partial}{\partial x}\left(
1-\frac{v_0^2}{c_s^2}\right)
\end{eqnarray}

Giovanazzi presented a microscopic description of Hawking radiation of
 sonic black holes  in Ref.\cite{Giovanazzi:2004zv}.
A one-dimensional Fermi-degenerate liquid squeezed by a smooth barrier
 $V=-\frac{1}{2}\lambda  x^2$ forms a transonic flow.
Using $p=\hbar^2\pi^2\rho^3/3m^4$ for the Fermi-degenerate liquid
and classical solutions
\begin{eqnarray}
\rho_0&=&\frac{m}{2\pi\hbar}\left(\sqrt{2m(E_F+\lambda x^2/2)}
-\sqrt{m\lambda}x
\right)\nonumber\\
v_0&=&\frac{1}{2m}\left(\sqrt{2m(E_F+\lambda x^2/2)}
+\sqrt{m\lambda}x
\right),
\end{eqnarray}
where $E_F$ is the energy of a particle on the Fermi surface,
one will see that the sound horizon appears at $x=0$.
Here the sound velocity is given by $c_s=\hbar\pi\rho_0/m^2$.
After quantization, one will find that the temperature of Hawking
radiation is given by
\begin{eqnarray}
T=\frac{\hbar}{2\pi k_B}\sqrt{\frac{\lambda}{m}}.
\end{eqnarray}
By choosing $\lambda=m\omega_x^2$, we have
\begin{eqnarray}
\label{eq-sonicT}
T=\frac{\hbar \omega_x}{2\pi k_B}.
\end{eqnarray}
Giovanazzi's microscopic treatment of the quantum process 
shows a close relationship between
 sonic Hawking radiation and quantum tunneling (and quantum scattering)
 through the barrier.

To be more specific about the quantum processes, 
particles with $E>0$ radiate positive energy via 
quantum scattering (back outward the horizon),
while particles with $E<0$
can tunnel the horizon barrier and enter the black hole.
These quantum processes seem to explain Hawking's original 
idea of ``Hawking radiation as tunneling''.
The temperature of the radiation given by Eq.(\ref{eq-sonicT})
will be calculated either using special functions or from 
$e^{K_{ud}}$ of the EWKB, since this problem 
is nothing more than scattering by an inverted quadratic potential.

On the other hand, a comparison of this model with our calculation of
the Stokes phenomena of the black hole shows a critical discrepancy
between them.

\section{How to include quantum backreactions in the EWKB}
\label{sec-qb}
To illustrate the quantum correction to the Stokes phenomenon of Hawking
radiation in the EWKB formalism, we first consider Hawking radiation
from the sonic black hole.
The result in this section can be easily applied to the Stokes phenomenon
of black hole radiation.
This is because in both cases the Stokes phenomena are explained by the MTP
structure of the inverted quadratic potential.
Given that Hawking radiation is described by the Stokes phenomenon of 
the local time coordinate system, the only parameter required for the
Stokes phenomenon is the local surface gravity.
Following Ref.\cite{Giovanazzi:2004zv}, we consider the EWKB
with a potential $V(x)=-\frac{1}{2}m a^2 x^2$, 
which gives the temperature of the sonic black hole
\begin{eqnarray}
T&=&\frac{\hbar}{2\pi k_B}a.
\end{eqnarray}
To compare the sonic black hole and the conventional Hawking radiation, 
we have relations
\begin{eqnarray}
a&\rightarrow& \frac{c^3}{4GM}\nonumber\\
\dot{M}&=&\frac{\hbar c^4}{15360 \pi G^2}\frac{1}{M^2},
\end{eqnarray}
where we set $G$ and $M$ are the gravitational constant and
the black hole mass, respectively.
We also have $\dot{a}\propto \dot{M}/{M^2}\propto M^{-4}\hbar\propto a^4\hbar$.

For a real black hole, the radiation continuously takes away the mass of
the black hole.
Then, what an observer sees as the black hole mass depends on the distance
from the horizon.
To realize the same phenomenon in the sonic black hole, we consider a
snapshot of the black hole and introduce
$a(x)\simeq a_0+ f(\hbar,x)$, where $a_0$ is defined
at the horizon ($x=0$).
Then the quantum correction can be expanded as
\begin{eqnarray}
f(x,\hbar)&=&\sum^\infty_{j=1} \hbar^j f_j(x).
\end{eqnarray}
Now the problem becomes scattering by a potential
\begin{eqnarray}
V(x,\hbar)&=&-\frac{m}{2}a_0^2x^2 -\frac{m}{2} \sum^\infty_{j=1} \hbar^j
 f_j(x).
\end{eqnarray}
For the EWKB, we have 
\begin{eqnarray}
Q(x)_0&=&-E-\frac{m}{2}a_0^2x^2,\nonumber\\
Q(x)_j&=&-\frac{m}{2} f_j(x),
\end{eqnarray}
where $j\ge 1$.
Then, what we have to do is (1) Draw the Stokes lines using $Q_0$
and (2) Calculate the $\hbar$-correction of the observables.
Since the Stokes lines are defined by $Q(x)_0$, the Stokes lines are
identical to the original model and
quantum corrections must appear only in $K_{ud}$.

\section{Conclusions and discussions}
In this paper, the EWKB analysis and the Stokes phenomena are used to 
analyze the Unruh effect and Hawking radiation.
The similarity between the Schwinger effect and the Unruh effect (Hawking
radiation) has often been mentioned in textbooks,
but there has been no clear explanation of the common Stokes phenomenon.
Also, the Unruh effect and Hawking radiation seem to be local
phenomena.
To address these issues, we focused on the Schwinger effect in a
constant electric field and examined its similarity to the Unruh effect
and Hawking radiation.
Writing down the modes of the inertial vacuum using the observer's time,
we found the Stokes phenomenon in the local time coordinate system.
The calculation gives the correct Hawking temperature.

We then compared our analysis of the black hole with an analog
black hole and found a crucial discrepancy between them. 

To show why the exact $\hbar$ expansion is important for
semiclassical calculations, we have briefly described how Hawking
 radiation is altered by the quantum radiation.

An application of these results is a bound on chaos \cite{Maldacena:2015waa}.
In order to investigate the chaos bound of a quantum system in detail,
it is useful to investigate the Stokes phenomenon causing the quantum 
radiation. 
At the same time, an accurate way to describe the relationship between
classical and quantum effects is needed \cite{Morita:2019bfr,Shudo:2016}. 
In this paper, the Stokes phenomenon of Hawking radiation is
analyzed with the EWKB, where the EWKB provides a way to describes the
relationship between the classical and quantum theoretical effects in an
accurate way, in the sense that its $\hbar$ expansion is rigorous.

\acknowledgments
The authors would like to thank Nobuhiro Maekawa for his collaboration in
the very early stages of this work.
SE  was supported by the Sun Yat-sen University Science Foundation.


\begin{thebibliography}{1}
\bibitem{Berry:1972na}
M.~V.~Berry and K.~Mount,
``Semiclassical approximations in wave mechanics,''
Rept.\ Prog.\ Phys.\  \textbf{35} (1972), 315
\bibitem{Landau-Lifshitz}
L.~D.~Landau and E.~M.~Lifshitz, ``Quantum Mechanics'' 
\bibitem{Parikh:1999mf}
M.~K.~Parikh and F.~Wilczek,
``Hawking radiation as tunneling,''
Phys. Rev. Lett. \textbf{85} (2000), 5042-5045
[arXiv:hep-th/9907001 [hep-th]].
\bibitem{Dumlu:2020wvd}
C.~K.~Dumlu,
``Stokes phenomenon and Hawking radiation,''
Phys. Rev. D \textbf{102} (2020) no.12, 125006
[arXiv:2009.09851 [hep-th]].
\bibitem{Akhmedova:2008dz}
V.~Akhmedova, T.~Pilling, A.~de Gill and D.~Singleton,
``Temporal contribution to gravitational WKB-like calculations,''
Phys. Lett. B \textbf{666} (2008), 269-271
[arXiv:0804.2289 [hep-th]].
\bibitem{Singleton:2013}
A.~de Gill, D.~Singleton, V.~Akhmedova and T.~Pilling,
``A WKB-Like Approach to Unruh Radiation,''
Am. J. Phys. \textbf{78} (2010), 685-691.
[arXiv:1001.4833 [gr-qc]].
\bibitem{Kraus:1994fj}
P.~Kraus and F.~Wilczek,
``Effect of selfinteraction on charged black hole radiance,''
Nucl. Phys. B \textbf{437} (1995), 231-242
[arXiv:hep-th/9411219 [hep-th]].
\bibitem{Keski-Vakkuri:1996wom}
E.~Keski-Vakkuri and P.~Kraus,
``Microcanonical D-branes and back reaction,''
Nucl. Phys. B \textbf{491} (1997), 249-262
[arXiv:hep-th/9610045 [hep-th]].
\bibitem{Srinivasan:1998ty}
K.~Srinivasan and T.~Padmanabhan,
``Particle production and complex path analysis,''
Phys. Rev. D \textbf{60} (1999), 024007
[arXiv:gr-qc/9812028 [gr-qc]].
\bibitem{Banerjee:2009wb}
R.~Banerjee and B.~R.~Majhi,
``Hawking black body spectrum from tunneling mechanism,''
Phys. Lett. B \textbf{675} (2009), 243-245
[arXiv:0903.0250 [hep-th]].
\bibitem{Shankaranarayanan:2000qv}
S.~Shankaranarayanan, T.~Padmanabhan and K.~Srinivasan,
``Hawking radiation in different coordinate settings: Complex paths approach,''
Class. Quant. Grav. \textbf{19} (2002), 2671-2688
[arXiv:gr-qc/0010042 [gr-qc]].
\bibitem{Arzano:2005rs}
M.~Arzano, A.~J.~M.~Medved and E.~C.~Vagenas,
``Hawking radiation as tunneling through the quantum horizon,''
JHEP \textbf{09} (2005), 037
[arXiv:hep-th/0505266 [hep-th]].
\bibitem{Martin:2002vn}
J.~Martin and D.~J.~Schwarz,
``WKB approximation for inflationary cosmological perturbations,''
Phys. Rev. D \textbf{67} (2003), 083512
[arXiv:astro-ph/0210090 [astro-ph]].
\bibitem{Iyer:1986np}
S.~Iyer and C.~M.~Will,
``Black Hole Normal Modes: A {WKB} Approach. 1. Foundations and Application of a Higher Order {WKB} Analysis of Potential Barrier Scattering,''
Phys. Rev. D \textbf{35} (1987), 3621
\bibitem{Konoplya:2003ii}
R.~A.~Konoplya,
``Quasinormal behavior of the d-dimensional Schwarzschild black hole and higher order WKB approach,''
Phys. Rev. D \textbf{68} (2003), 024018
[arXiv:gr-qc/0303052 [gr-qc]].
\bibitem{Konoplya:2011qq}
R.~A.~Konoplya and A.~Zhidenko,
``Quasinormal modes of black holes: From astrophysics to string theory,''
Rev. Mod. Phys. \textbf{83} (2011), 793-836
[arXiv:1102.4014 [gr-qc]].
\bibitem{NBR}
H.~Berk,W.~Nevins and K.~Roberts,``New Stokes line in WKB
	theory''. Journal of Mathematical Physics, 23 (1982) 988-1002.
\bibitem{Virtual:2015HKT}
N.~Honda, T.~Kawai and Y.~Takei,
``Virtual Turning Points'', Springer (2015),
ISBN 978-4-431-55702-9.
\bibitem{RPN:2017}
``Resurgence, Physics and Numbers'' edited by F.~Fauvet, D.~Manchon,
	S.~Marmi and  D.~Sauzin, Publications of the Scuola Normale
	Superiore, 978-88-7642-613-1
\bibitem{Delabaere:1993}
E. Delabaere, H. Dillinger and F. Pham: Resurgence de Voros et peeriodes
des courves hyperelliptique. Annales de l'Institut Fourier, 43 (1993), 163-
199.
\bibitem{Silverstone:2008}
H.~Shen and H.~J.~Silverstone, ``Observations on the JWKB treatment of
the quadratic barrier, Algebraic analysis of differential equations from
	microlocal analysis to exponential asymptotics'', Springer,
	2008, pp. 237 - 250.
\bibitem{Takei:2008}
Y.~Takei, ``Sato's conjecture for the Weber equation and
transformation theory for Schrodinger equations
with a merging pair of turning points'',
RIMS Kokyuroku Bessatsu B10 (2008), 205-224,
\begin{verbatim}
https://www.kurims.kyoto-u.ac.jp/~kenkyubu/bessatsu/open/B10/pdf/B10_011.pdf
\end{verbatim}
\bibitem{Aoki:2009}
T.~Aoki, T.~Kawai, and T.~Takei, ``The Bender-Wu analysis and the Voros
theory, II'', Adv. Stud. Pure Math. 54 (2009), Math. Soc. Japan, Tokyo, 2009,
pp. 19 -94
\bibitem{Voros:1983}
A.~Voros, ``The return of the quartic oscillator -- The complex WKB
method'', Ann. Inst. Henri Poincare, 39 (1983), 211-338.
\bibitem{Kofman:1997yn}
  L.~Kofman, A.~D.~Linde and A.~A.~Starobinsky,
  ``Towards the theory of reheating after inflation,''
  Phys.\ Rev.\ D {\bf 56} (1997) 3258
  [hep-ph/9704452].
\bibitem{Enomoto:2020xlf}
S.~Enomoto and T.~Matsuda,
``The exact WKB for cosmological particle production,''
JHEP \textbf{03} (2021), 090
[arXiv:2010.14835 [hep-ph]].
\bibitem{Zener:1932ws}
  C.~Zener,
  ``Nonadiabatic crossing of energy levels,''
  Proc.\ Roy.\ Soc.\ Lond.\ A {\bf 137} (1932) 696.
\bibitem{Enomoto:2021hfv}
S.~Enomoto and T.~Matsuda,
``The exact WKB and the Landau-Zener transition for asymmetry in cosmological particle production,''
JHEP \textbf{02} (2022), 131
[arXiv:2104.02312 [hep-th]].
\bibitem{WKB-recent:2022}
S.~K.~Ashok, D.~P.~Jatkar, R.~R.~John, M.~Raman and J.~Troost,
``Exact WKB analysis of $ \mathcal{N} $ = 2 gauge theories,''
JHEP \textbf{07} (2016), 115
[arXiv:1604.05520 [hep-th]];
F.~Yan,
``Exact WKB and the quantum Seiberg-Witten curve for 4d N = 2 pure SU(3) Yang-Mills. Abelianization,''
JHEP \textbf{03} (2022), 164
[arXiv:2012.15658 [hep-th]];
A.~Grassi, Q.~Hao and A.~Neitzke,
``Exact WKB methods in SU(2) N$_{f}$ = 1,''
JHEP \textbf{01} (2022), 046
[arXiv:2105.03777 [hep-th]];
S.~Kamata, T.~Misumi, N.~Sueishi and M.~\"Unsal,
``Exact-WKB analysis for SUSY and quantum deformed potentials: Quantum mechanics with Grassmann fields and Wess-Zumino terms,''
[arXiv:2111.05922 [hep-th]];
M.~Alim, L.~Hollands and I.~Tulli,
``Quantum curves, resurgence and exact WKB,''
[arXiv:2203.08249 [hep-th]];
M.~F.~Girard,
``Exact WKB-like Formulae for the Energies by means of the Quantum Hamilton-Jacobi Equation,''
[arXiv:2204.02708 [quant-ph]];
A.~van Spaendonck and M.~Vonk,
``Painlev\'e I and exact WKB: Stokes phenomenon for two-parameter transseries,''
[arXiv:2204.09062 [hep-th]];
K.~Imaizumi,
``Quasi-normal modes for the D3-branes and Exact WKB analysis,''
Phys. Lett. B \textbf{834} (2022), 137450
[arXiv:2207.09961 [hep-th]].
\bibitem{Unruh:1976db}
W.~G.~Unruh,
``Notes on black hole evaporation,''
Phys. Rev. D \textbf{14} (1976), 870
\bibitem{Hawking:1975vcx}
S.~W.~Hawking,
``Particle Creation by Black Holes,''
Commun. Math. Phys. \textbf{43} (1975), 199-220
[erratum: Commun. Math. Phys. \textbf{46} (1976), 206]
\bibitem{Enomoto:2022nuj}
S.~Enomoto and T.~Matsuda,
``The Exact WKB analysis for asymmetric scalar preheating,''
[arXiv:2203.04497 [hep-th]].
\bibitem{Koike:2000}
T.~Koike, ``On the Exact WKB Analysis of Second Order Linear Ordinary
	Differential Equations with Simple Poles'', Publications of the
	Research Institute for Mathematical Sciences, \textbf{36}
	(2000), 297-319.
\bibitem{Maldacena:1997ih}
J.~M.~Maldacena and A.~Strominger,
``Universal low-energy dynamics for rotating black holes,''
Phys. Rev. D \textbf{56} (1997), 4975-4983
[arXiv:hep-th/9702015 [hep-th]].
\bibitem{Birrell:1982ix}
N.~D.~Birrell and P.~C.~W.~Davies,
``Quantum Fields in Curved Space,''
\bibitem{Frolov:1998wf}
V.~P.~Frolov and I.~D.~Novikov,
``Black hole physics: Basic concepts and new developments,''
\bibitem{Aokisteepest:2004}
T.~Aoki, T.~Kawai, and Y.~Takei, ``The exact steepest descent method - a
	new steepest descent method based on the exact WKB analysis'',
 Adv. Stud. Pure Math. \textbf{42}  (2004), 45.
\bibitem{Chung:1998bt}
D.~J.~Chung,
``Classical Inflation Field Induced Creation of Superheavy Dark Matter,''
Phys.\ Rev.\ D \textbf{67} (2003), 083514
[arXiv:hep-ph/9809489 [hep-ph]].
\bibitem{Enomoto:2013mla}
S.~Enomoto, S.~Iida, N.~Maekawa and T.~Matsuda,
``Beauty is more attractive: particle production and moduli trapping with higher dimensional interaction,''
JHEP \textbf{01} (2014), 141
[arXiv:1310.4751 [hep-ph]].
\bibitem{Enomoto:2014hza}
S.~Enomoto, N.~Maekawa and T.~Matsuda,
``Preheating with higher dimensional interaction,''
Phys.\ Rev.\ D \textbf{91} (2015) no.10, 103504
[arXiv:1405.3012 [hep-ph]].
\bibitem{Dabrowski:2016tsx}
R.~Dabrowski and G.~V.~Dunne,
``Time dependence of adiabatic particle number,''
Phys. Rev. D \textbf{94} (2016) no.6, 065005
[arXiv:1606.00902 [hep-th]].
\bibitem{Haro:2010mx}
J.~Haro,
``Topics in Quantum Field Theory in Curved Space,''
[arXiv:1011.4772 [gr-qc]].
\bibitem{Schwinger:1951nm}
J.~S.~Schwinger,
``On gauge invariance and vacuum polarization,''
Phys. Rev. \textbf{82} (1951), 664-679
\bibitem{Giovanazzi:2004zv}
S.~Giovanazzi,
``Hawking radiation in sonic black holes,''
Phys. Rev. Lett. \textbf{94} (2005), 061302
[arXiv:physics/0411064 [physics]].
\bibitem{Aoki:1993}
T.~Aoki, J.~Yoshida, ``Microlocal Reduction of Ordinary
	Differential Operators with a Large Parameter'', 
Publications of the Research Institute for Mathematical Sciences,
	\textbf{29} (1993), 959-975
\bibitem{Aoki:2019}
Takashi~Aoki, Kohei~Iwaki, Toshinori~Takahashi, ``Exact WKB Analysis of
	Schrodinger Equations with a Stokes Curve of Loop Type'',
	Funkcialaj Ekvacioj, 2019, Volume 62, Issue 1, Pages 1-34 
\bibitem{Taya:2020dco}
H.~Taya, T.~Fujimori, T.~Misumi, M.~Nitta and N.~Sakai,
``Exact WKB analysis of the vacuum pair production by time-dependent electric fields,''
JHEP \textbf{03} (2021), 082
[arXiv:2010.16080 [hep-th]].
\bibitem{Kitamoto:2020tjm}
H.~Kitamoto,
``No-go theorem of anisotropic inflation via Schwinger mechanism,''
[arXiv:2010.10388 [hep-th]].
\bibitem{Martin:2007bw}
J.~Martin,
``Inflationary perturbations: The Cosmological Schwinger effect,''
Lect. Notes Phys. \textbf{738} (2008), 193-241
[arXiv:0704.3540 [hep-th]].
\bibitem{Dumlu:2010ua}
C.~K.~Dumlu and G.~V.~Dunne,
``The Stokes Phenomenon and Schwinger Vacuum Pair Production in Time-Dependent Laser Pulses,''
Phys. Rev. Lett. \textbf{104} (2010), 250402
[arXiv:1004.2509 [hep-th]].
\bibitem{Nakamura:2007vp}
T.~K.~Nakamura,
``Factor two discrepancy of Hawking radiation temperature,''
[arXiv:0706.2916 [hep-th]].
\bibitem{Maldacena:2015waa}
J.~Maldacena, S.~H.~Shenker and D.~Stanford,
``A bound on chaos,''
JHEP \textbf{08} (2016), 106
[arXiv:1503.01409 [hep-th]].
\bibitem{Morita:2019bfr}
T.~Morita,
``Thermal Emission from Semi-classical Dynamical Systems,''
Phys. Rev. Lett. \textbf{122} (2019) no.10, 101603
[arXiv:1902.06940 [hep-th]].
\bibitem{Shudo:2016}
A.~Shudo and K.~S.~Ikeda
``Toward pruning theory of the Stokes geometry for the quantum Henon
	map''
Nonlinearity \textbf{29} (2016) 375
\end{thebibliography}
\end{document}